# The origin of the genetic code is encrypted in the structure of present-day transfer RNAs

**J. H. Daniel**

*InvenTsion, Rehovot, Israel*

jacques.daniel1@gmail.com

**Background/Objectives:** Resolving the origin of the genetic code is fundamental to understanding how Life began its journey out of the chemical world. Since its deciphering some 60 years ago, there is still no general theory of the emergence of the genetic code. My objectives are to bring some unique data that might provide some insight into this particular issue. **Methods:** Because tRNA (transfer RNA) constitutes a crucial piece of the present translational system, having unique structural characteristics, I hypothesized that they might constitute the key elements at the origin of the genetic code, and thus, decided to compare the primary structure of the tRNAs from a bacterium, *Bacillus subtilis*.**Results:** The comparison of the primary structure of the tRNAs from *Bacillus subtilis*, generated a genealogical tree, meaning that the tRNAs were all related and appeared gradually in a precise time sequence. Remarkably, analysis of the various characteristics of this tRNAs tree showed that it very likely reflects the time of entry of amino acids into the Universal Codon Table. **Conclusions:** These results strongly suggest that the tRNA entity was indeed a major component in the formation of the genetic code, and further, provide a likely scenario for the time sequence of codon colonization of the Universal Codon Table by the various amino acids at the very beginning of life. Also, these data are interpreted in terms of a general theory of the origin of the genetic code I propose, the poly-tRNA theory.

1. Introduction

Genetic coding –a formidable biological phenomenon thought to have successfully appeared only once and this, some billions of years ago –certainly has a very complex origin which has remained largely unresolved, despite all of the fantastic advances of biochemistry, molecular biology, genetics and genomics these last decades. Several theories related to the origin of the genetic code exist (see [1, 2]), such as the Frozen Accident Theory proposed by Francis Crick [3], the Co-evolution Theory [4], the Stereo-chemical Theory [5,6], and the Error Minimization Theory [7]. Although most of these theories bring interesting observations and correlations, they are generally rather vague, non-specific and uncommitted as to how genetic coding might have occurred. This indecision and ultimately, failure to enlighten us on the possible steps and mechanisms that might have given rise to the genetic code were my primary motivation to dig into this fundamental issue. In a previous work, inspired by the profound intuition of Crick that "it is almost impossible to discuss the origin of the code without discussing the origin of the actual biochemical mechanisms of protein synthesis" [3], an original model was proposed for the origin and evolution of genetic coding [8,9],. Here, I wish to extend this study and focus more specifically on the likely time sequence of amino acids' colonization of the Universal Codon Table, as revealed by tRNA nucleotide sequence comparisons, and see how these data could suggest successive plots for the origin of genetic coding.

2. Materials and Methods

The program used for the comparison of the various tRNAs from *Bacillus subtilis* was LALIGN (see [8]). For values of Similarity Index and way to compute them, see Fig2 A, B and legend in [8].

3. Results and Discussion

*Time sequence of codon adoption by amino acids*

As previously shown [8], it is remarkable that the tRNAs from the gram-positive model bacterium *Bacillus subtilis* (completed by tRNAs from the gram-negative *Escherichia coli*; two representatives of the bacterial reign which is possibly the oldest one in the living world) are all related to each other by their nucleotide sequences, forming a genealogical tree with two main branches starting with valine and tyrosine, respectively (Fig. 1). This tree would have been generated by duplication and mutation from the direct ancestor of each of its components, all of the components remaining together inside the proto-cell. Noticeably, most tRNAs in Box 1 (i.e. the most ancient) correspond to those that today use the class II amino acyl-tRNA synthetases, except for tRNA-Leu and tRNA-Ile. It was proposed that the tRNAs binding leucine, isoleucine and valine were originally processed by class II amino acyl-tRNA synthetases before shifting to class I amino acyl-tRNA synthetases [9], the latter believed to have appeared more recently evolutionarily and thus, possibly having better editing capacity [10]. This is indeed a reasonable hypothesis considering proteins' difficulty in discriminating between these relatively small hydrophobic amino acids. Conversely, in Box 2 (the most recent tRNAs) six out of seven tRNAs are processed by the likely more recent class I amino acyl-tRNA synthetases.

Another striking feature in this genealogical tree is that all seven tRNAs of Box 2 belong to the square tiles (defined by the two first codon letters) of the Universal Codon Table that are *shared* with a tRNA corresponding to another amino acid (there are, altogether, 7 shared square tiles out of the 16 square tiles composing the Table) (Fig. 1 and Fig. 4 for the Universal Codon Table, in background).

These findings greatly strengthen the conclusion that the tRNA genealogical tree displayed here might be an essential and fascinating attribute for understanding crucial aspects of the origin of genetic coding. Indeed, on the basis of the three mentioned criteria: (i) box assignation, i.e., type of amino acyl-tRNA synthetase used, (ii) order of appearance of the two classes of amino acyl-tRNA synthetase in the tree, and (iii) type of tile used in the Universal Codon Table, the probability of observing this tree by chance is only about one in a million, provided we change the synthetase type for valine, leucine and isoleucine, according to the hypothesis stated above; by not including this provision, the probability is somewhat less, but still very highly significant.

**Figure 1:**

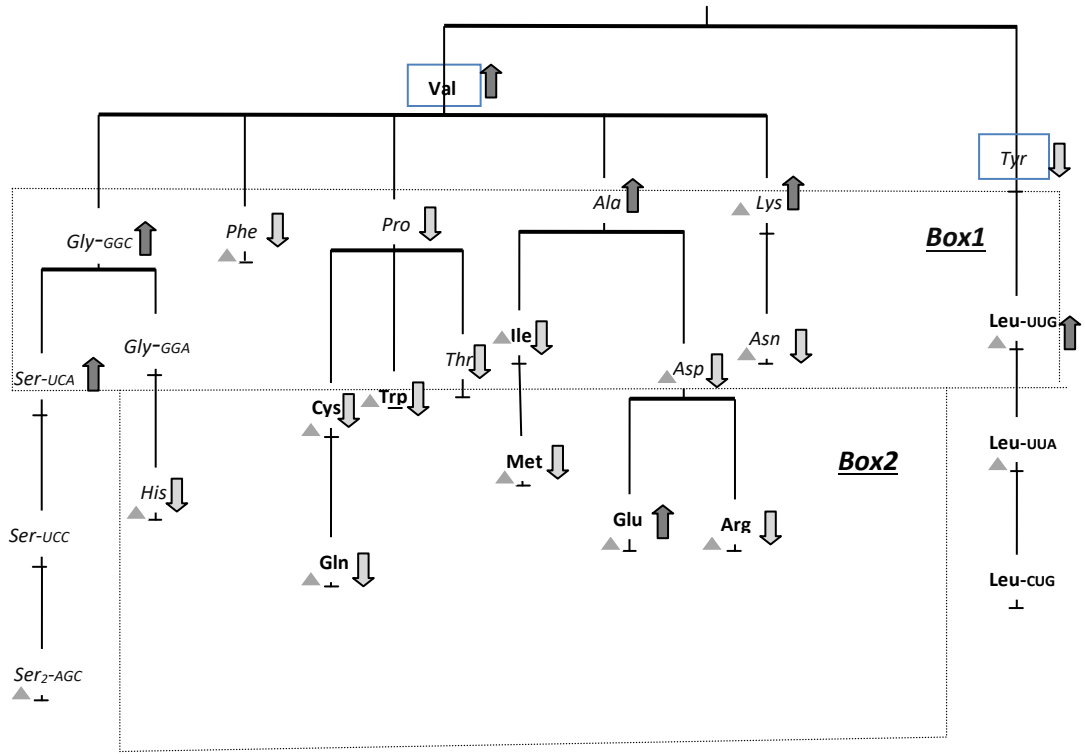

**Fig. 1**: **Genealogical tree of all tRNAs from *Bacillus subtilis* obtained by nucleotide sequence comparison**. For the construction of the tree, see Daniel [8]. At each node of the tree, tRNAs are represented by their corresponding amino acids (as three-letter symbols). Grey amino acids in italics: tRNAs using class II amino acyl-tRNA synthetases. Black amino acids in bold: tRNA using class I amino acyl-tRNA synthetases. Triangles represent amino acids whose corresponding codons belong to shared square tiles in the Universal Codon Table. Upward black arrows are for amino acids over-represented in proteins; downward grey arrows, amino acids under-represented in proteins. The potentially most accurate average of amino acid proportion in today's proteins (in %) was obtained by AI search, as follows (from smallest to largest):  Trp (tryptophan) 1.3/ Cys (cysteine) 1.5/ His (histidine) 2.1/ Met (methionine) 2.4/ Tyr (tyrosine) 3.2/ Phe (phenylalanine) 3.8/ Gln (glutamine) 3.9/ Asn (asparagine) 4.4/ Ile (isoleucine) 4.5/ Pro (proline) 4.7/ Arg (arginine) 5.1/ Asp (aspartic acid) 5.3/ Thr (threonine) 5.4/ Lys (lysine) 5.9/ Glu (glutamic acid) 6.2/ Val (valine) 6.7/ Ser (serine) 6.9/ Gly (glycine) 7.1/ Ala (alanine) 8.3/ Leu (leucine) 9.0. The limit between over-expression and under-expression in proteins was taken as 5.5%. See text.

As shown in Fig. 2 A-D, the tRNAs [amino acids] were separated in four groups according to their relative time of arrival during prebiotic evolution. Represented for each time phase, is (i) the codon recognized by the specific tRNA-amino acid as positioned in the Universal Codon Table, (ii) the nature of the codon, written to emphasize the first and second nucleotides, and (iii) the arrival time of the specific tRNA-amino acid relative to time 0, corresponding to the putative common ancestor of all of these tRNAs.

The first group (Fig 2 A) includes valine, and then after some time period, tyrosine (each at the origin of one of the two main branches issued from their putative common ancestor). The next two groups belong to Box1 of Fig. 2, and consist of two waves of tRNA creation (Fig 2 B, C), each wave comprising several amino acids having emerged more or less synchronously. The last group refers to a rather large set of new tRNAs which appeared progressively within a relatively long period of time (Fig 2 D). Also shown in Fig 2 D, is the average relative "speed" of tRNA creation for each of the four time phases, represented by a dashed-dotted line whose slope, relative to the horizontal line, is inversely proportional to this speed.

**Figure 2:**

A

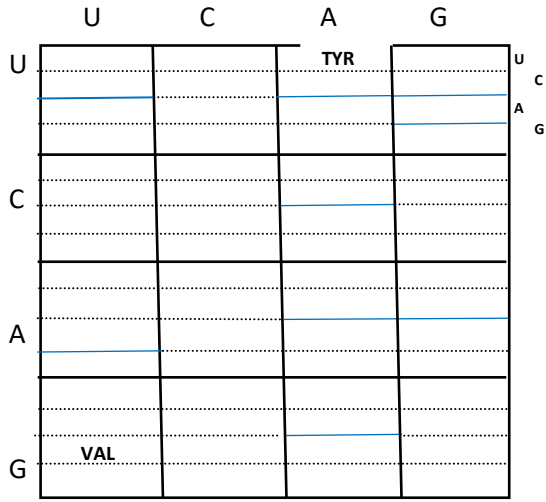
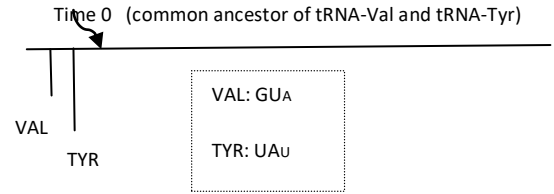

*Box1-a:*

B

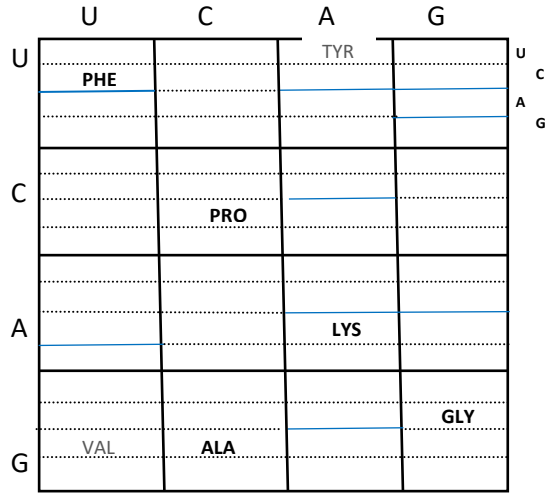
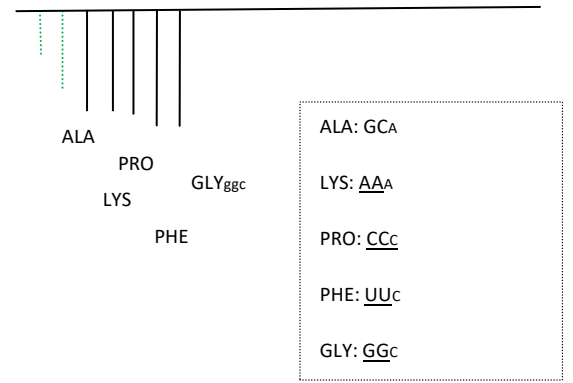

**Time** ↓

**C**

*Box1-b:*

|   | U | C | A | G |   |
|---|---|---|---|---|---|
| U | PHE | SER | TYR |  | U<br>C<br>A<br>G |
|   | **LEU** |  |  |  |   |
| C |  | PRO |  |  |   |
| A | **ILE** | **THR** | **ASN** |  |   |
|   |  |  | LYS |  |   |
| G | VAL | ALA | ASP | GLY |   |
|   |  |  |  | **GLY** |   |

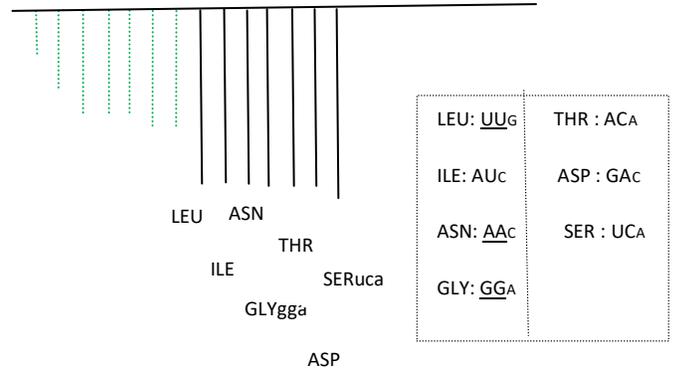

LEU   ASN
         THR
ILE        SERuca
      GLYgga
            ASP

| LEU: **UU**G | THR : ACA |
| ILE: AUC | ASP : GAC |
| ASN: **AA**C | SER : UCA |
| GLY: **GG**A |  |

**Time** ↓

**D**

*Box2 (plus LEU, SER ; and from E.coli):*

|   | U | C | A | G |   |
|---|---|---|---|---|---|
| U | PHE | SER | TYR | CYS | U<br>C<br>A<br>G |
|   | **LEU** | SER |  |  |   |
|   | LEU | + |  | **TRP** |   |
| C | + | + | HIS | ARG |   |
|   | + | PRO | GLN |  |   |
|   | **LEU** | + | + | + |   |
| A | ILE | + | ASN | **SER** |   |
|   |  | **THR** | LYS | + |   |
|   | **MET** | + |  | + |   |
| G | + | + | ASP | GLY |   |
|   | VAL | ALA | **GLU** | GLY |   |

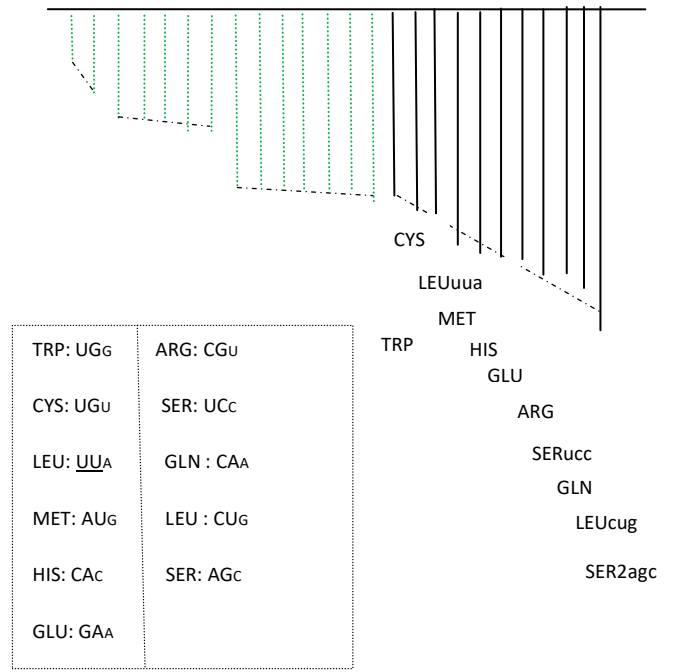

                CYS
              LEUuua
                MET
            TRP   HIS
                    GLU
                       ARG
                         SERucc
                           GLN
                             LEUcug
                               SER2agc

| TRP: UGG | ARG: CGU |
| CYS: UGU | SER: UCC |
| LEU: **UU**A | GLN : CAA |
| MET: AUG | LEU : CUG |
| HIS: CAC | SER: AGC |
| GLU: GAA |  |

**Fig. 2**: **Time sequence of colonization of the Universal Codon Table by amino acids**. A snapshot of codon adoption by amino acids is given for each of the four defined time phases (from **A** to **D**). Underlined bases of codons represent their identical first and second bases (i.e., codons belonging to one diagonal of the Universal Codon Table). The horizontal line of the graph defines time zero, corresponding to the common ancestor of valine and tyrosine tRNA. The length of each vertical line drawn from this time zero line indicates the relative time needed for the appearance of the specific tRNA-[amino acid]. Dashed-dotted line in **D** represents the relative speed of tRNA-amino acid appearance for each time phase (see text). In **D**, the + symbol indicates additional data obtained from *Escherichia coli*. See Daniel [8].

*Remarkable positions of amino acids in the Universal Codon Table*

It is very significant that, as mentioned above, all seven tRNAs of Box 2 are found in shared square tiles of the Universal Codon Table. The simplest and most straightforward explanation is that their late arrival in primal evolution would have left them only with the possibility to occupy the remaining free codons, or alternatively albeit less likely, to push away and then replace the assignation of some codons already occupied by another amino acid such as lysine, isoleucine or aspartate, for example (see Universal Codon Table as background of Fig 4).

Moreover, it is intriguing that, included in the sequence of tRNAs (12 in total) formed by the two waves of Box 1 together, there are 7 tRNAs corresponding to codons made up of identical first and second nucleotides (underlined in fig 2 B, C). Even more striking, in the first wave of Box1, four out of the five tRNAs present this peculiar feature of their corresponding codons. Since the probability of getting the last result by chance alone is less than 0.004, in the following, I try to make some possible sense of these unexpected findings.

*Codon position of amino acids and their average proportion in today's proteins*

The average proportion of each of the 20 amino acids composing proteins varies relatively widely, between 1.3% for the lowest (tryptophan), and 9% for the highest (leucine), these values being rather far from the 5% value expected if all amino acids were equally represented in proteins (Fig. 1). If we take the value of 5.5% as the separation limit between the under-represented and over-represented amino acids, we notice that in Box 1, to which valine and tyrosine are added, there are similar amounts of under-represented and over-represented amino acids (7 vs. 6, respectively). However, in Box 2, six out seven amino acids are under-represented (p = 0.015; average 2.7%), glutamate being the only amino acid that is over-represented. This is interesting because it strongly suggests that deficit of over-represented amino acids in Box 2 is not due to these amino acids' recent arrival in prebiotic evolution or their belonging to *shared* square tiles of the Universal Codon Table. Rather, the most likely interpretation is that the under-represented amino acids, making up the great majority of amino acids in Box 2, are refinement and improvement increments in general protein structure, function and interaction, whereas the only real fundamental innovation at this phase of the code's development would concern glutamate.

*Putative mechanisms at the origin of genetic coding*

Results presented here clearly point to the fact that tRNA assumed a major role in establishing the universal genetic coding system of the living world, and this, likely at an early stage. These findings also strongly suggest that the processes involved in such a formidable construction took some evolutionary time and were put in place progressively. Indeed, at least four different time phases can be defined, each with its own characteristics, one of them being the relative speed with which the tRNA-amino acids appeared (Figs. 1 and 2).

A theory of the origin of the genetic code was previously proposed that is essentially based on tRNA as the main agent of the whole plot, initially acting in a putative polymeric form (poly-tRNA) [8]. The present and prior findings seem to fit well with– but of course, do not prove –such a theory.

For the two amino acids valine, and later tyrosine, constituting the first phase in the genetic coding process, the poly-tRNA theory suggests that the corresponding starting tRNA's motif contained, at its 5' end, a trinucleotide sequence capable of binding the amino acid, and that an identical sequence would also be found on the "anticoding" side of this tRNA motif (the possibility of such an event has been discussed previously [8]). To increase the probability of inducing an amide bond between amino acids attached to two contiguous tRNA motifs on the poly-tRNA, it was proposed that an RNA molecule with a continuous sequence made of consecutive subsequences that are complementary to each tRNA trinucleotide, would bind to the tRNA "anticoding" sides. That the binding of such an RNA molecule at the "anticoding "sides of two tRNA motifs might indeed facilitate (by some kind of primordial catalytic effect) the formation of the amide bond between the amino acids borne at their 5' side, could be supported by the finding that the two sides (anticoding and 5' end) of free tRNAs move in an anti-correlated way due to a hinge situated at their elbow [11]. Furthermore, it was suggested that later in primal evolution, these helper RNAs would be at the origin of the mRNAs (messenger RNAs) [8].

In the second phase of the genetic coding process concerning five additional amino acids, the finding that four of them correspond to codons that are each made of identical first and second nucleotides (one diagonal in the Universal Codon Table; Fig 2 B) is, as already noted, highly significant. This significance seems to be reinforced by the fact in the codons' genealogical tree (pairing the"anticodon" sides of tRNAs), these tRNA-amino acids (second phase) show a relationship with the tRNA-amino acid of the first phase (i.e., valine) that is only at the limit of significance. In contrast, significance in both the second phase vs. third phase and third vs. fourth phase is very high (Fig. 3). These results suggest the existence of some kind of driving force (natural selection) at the origin of the special type of codons found for these amino acids. I propose that the purpose of the double identical nucleotides at codon positions 1 and 2 is to strengthen the binding of the tRNA to the helper RNA postulated by the poly-tRNA theory, by adding a stacking effect and thus, increasing the efficiency of the amide reaction between adjacent amino acids at this early stage of genetic coding. As to alanine, the fifth amino acid of phase 2, the first two bases of its codon are GC, a base combination that is also known for its strong stacking effect.

**Figure 3:**

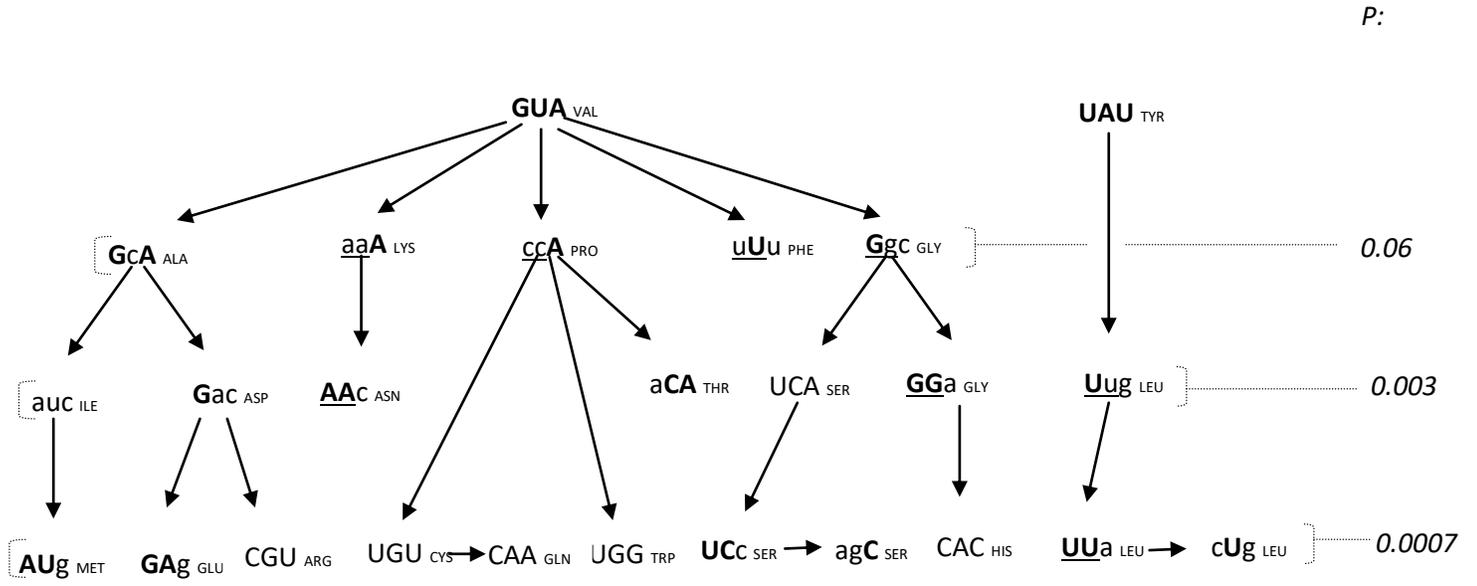

**Fig. 3: Homology between codons corresponding to the tRNAs arranged in time phases according to the genealogical tree**. The last phase includes all of the tRNAs appearing after Box 1-b. Total probability of occurrence by chance for each phase is given on the right. Underlined bases represent identical first and second codons. See text.

As one critical feature of the poly-tRNA theory, it is assumed that at some stage of the genetic code's creation, isolated tRNAs appeared and coexisted with poly-tRNAs, competing with them in peptide/protein production with identical RNA helpers, before the complete elimination of poly-tRNAs at a later stage [8]. It is interesting to attempt to somehow localize this crucial event on the tRNA genealogical tree presented here. Since the synthetases would likely be related to *isolated* tRNA (their function today) and there is a clear divide in the tree between the two types of synthetase used (i.e., the separation between Box 1 and Box 2), we can postulate that isolated tRNAs had already appeared before Box 2 and so, possibly in Box 1, because it seems difficult to consider that this could have already occurred at the very first valine/tyrosine stage. Indeed, it is likely that utilization of isolated tRNAs would require– even with the likely assistance of RNA tertiary structures –the presence of more complex peptides than those containing the sole valine and tyrosine residues, for example as part of proto-ribosomes, amino acid activation, proper synthetase reaction and so on. Tentatively, tRNA individuation might have occurred after the first boost of tRNA-amino acid creation at the beginning of Box 1, and continued with further refinement till the second boost of Box 1, or just after it.

Thus, we might envision the start and consolidation of the genetic coding process as follows. First, a tRNA motif binding valine and later, another one binding tyrosine, both being part of poly-tRNA structures, would set the stage for, and somehow prime, the nascent coding system, by essentially initiating the amide reaction between those amino acids, and using pairing with helper RNAs and possibly some other RNA tertiary structures to increase the efficiency of this reaction. At the second stage, and within a relatively short time period, five additional amino acid- binding motifs would be recruited, creating a much larger palette of amino acid residues for the production of more differentiated peptides/proteins, thereby opening up all sorts of new possibilities such as catabolic and anabolic metabolism, and structural and functional improvements of the proto-cell system, among others. Importantly, it would induce a remarkable virtuous spiral that would considerably improve the amide reaction's efficiency by starting the construction of a proto-ribosome structure. At this stage as well, isolated tRNAs would appear that would compete with poly-tRNA for binding to helper RNAs [8] (see earlier). The third stage would enable the protein-synthesizing machinery to incorporate more varied amino acid residues in the ever-richer panoply of proteins. In fact, the introduction of serine and threonine, both uncharged polar amino acid residues, could have contributed to protein-folding and stability, as well as enabling some primitive enzyme regulation, or even cellular signaling. In addition, the arrival of the first acidic amino acid, aspartate, would enhance protein solubility, allow ionic protein-protein interactions, and be involved in enzymes' active site for the catalytic reaction or substrate binding. Finally, at the last stage, the introduction of seven new amino acids would further refine all of the present and future proteins in their structure, function and interaction.

Concerning the emergence of isoacceptor tRNAs, there seems to be some correlation between their presence/number and the general protein content of their respective amino acid (for both *Bacillus subtilis* and *Escherichia coli*); thus, the capacity to derive isoacceptor tRNAs might constitute a kind of power reserve at the proto-cell's disposal. Although these tRNAs are rather closely related to their isoacceptor ancestor as well as to each other, it is remarkable that they have nevertheless accumulated mutations at sites other than the sole "anticodon" region. This might be considered divergent evolution, but with the same purpose of incorporating identical amino acids during protein synthesis. This finding seems to strongly suggest that the clock acting behind the time of tRNA appearances in the genealogical tree is essentially made up of neutral mutations relative to tRNA function [12, 13], thus confirming that "there exists a clear degree of plasticity in the canonical structure of the tRNA core that acts to buffer against the tertiary level effects of sequence variation and nucleobase modifications" as proposed by Chan *et al*. [14].

*Relevance of previous theories on the genetic code*

In view of the above analysis of the tRNA genealogical tree, it is interesting to evaluate the contribution brought, in a rather general way, by the previous theories on the genetic code. First, there is no convincing evidence in favor of the co-evolution theory [4]. As a matter of fact, each of the 13 amino acids first recruited in the tree is an amino acid consistently identified in prebiotic chemistry experiments as well as in meteorites, and appears to find its entry in the Universal Codon Table rather readily, with no obvious sign of displacement of another amino acid that would be its metabolic precursor.

In contrast, the stereo-chemical theory [5, 6] might be relevant for understanding the code-building process, as proposed herein and previously [8]. Conceivably, according to the poly-tRNA theory, valine and tyrosine, as well as the amino acids of Box 1-a, would bind to a specific site at their corresponding tRNA module (as part of poly-tRNA) that would also dictate the nature of their "anticodon" site [8, 15]. The possible effect of peptides made up of stretches of valine or valine/tyrosine on the membrane of proto-cells has been previously discussed [9] (also [16]). Adding to these two first amino acids the others from Box1-a would allow the synthesis of a whole panoply of peptides/proteins that would propel the proto-cell system to a completely different level. Indeed, apart from boosting primal metabolic pathways, it would also enable the priming of a virtuous spiral that would result in the continuous refinement of what would become the protein-synthesis machinery as we know it today. In particular, it would initiate the construction of the proto-ribosome and following the first appearance of isolated tRNAs at some stage later, it would create the primal amino acyl-tRNA synthetases. Moreover, the helper RNAs –the putative ancestors of mRNAs –would expand dramatically in length as well as variety.

An observation made early after the deciphering of the genetic code was that the Universal Codon Table is rather robust and resistant to mutational changes or read out errors. Although this feature does not seem to be the main clue to the origin of its formation –because one can demonstrate that there are billions of possible configurations more robust than the Universal Codon Table [2, 17] – it is interesting to consider what might have contributed to this desirable property. The first amino acid appearing in the tRNA genealogical tree is valine, a hydrophobic amino acid positioned in the bottom left-hand corner of the Universal Codon Table, as shown in Fig. 4. Valine will originate, as a first generation, three other hydrophobic amino acids: phenylalanine in the upper left corner (i.e., same second base as valine), alanine (valine's close neighbour, with the same first base) and glycine (in the bottom right corner, same first base as valine). In addition, two more amino acids will be directly generated from valine: one – proline –hydrophobic, and the other –lysine –hydrophilic, both positioned in the remaining square tiles of the diagonal defined by all of the codons having identical first and second bases. As a second generation, alanine will produce isoleucine, another hydrophobic amino acid, with the same second base as valine; and isoleucine will then generate the hydrophobic methionine, both belonging to the same square tile. As to the second branch of the tRNA genealogical tree –the one starting with tyrosine –we find only one type of amino acid generated, and this, rather late, i.e., the hydrophobic leucine. If we suppose that this direct descendant of tyrosine was more likely to colonize square tiles of the first two (identical) bases on the diagonal (as strongly suggested by the descendants of valine) and also retain either the first or second base of the tyrosine codon, only two choices would have been allowed: the phenylanine or lysine square tile. The fact that leucine found its way to the phenylalanine square tile –with only hydrophobic acids appearing in the first column of the codon table –might have resulted from natural selection for code robustness. Similar natural selection could have also operated for the descendant of lysine, i.e., asparagine, and that of aspartate, i.e., glutamate, each descendant sharing the same square tile as its immediate ancestor. Thus, simple rules, such as direct descendance by frequent conservation of the first or second base of the codon (or both) (see the notion of neutral emergence in [18, 19]), as well as the force of natural selection, might be at the origin of the rather robust Universal Codon Table we know today.

**Figure 4:**

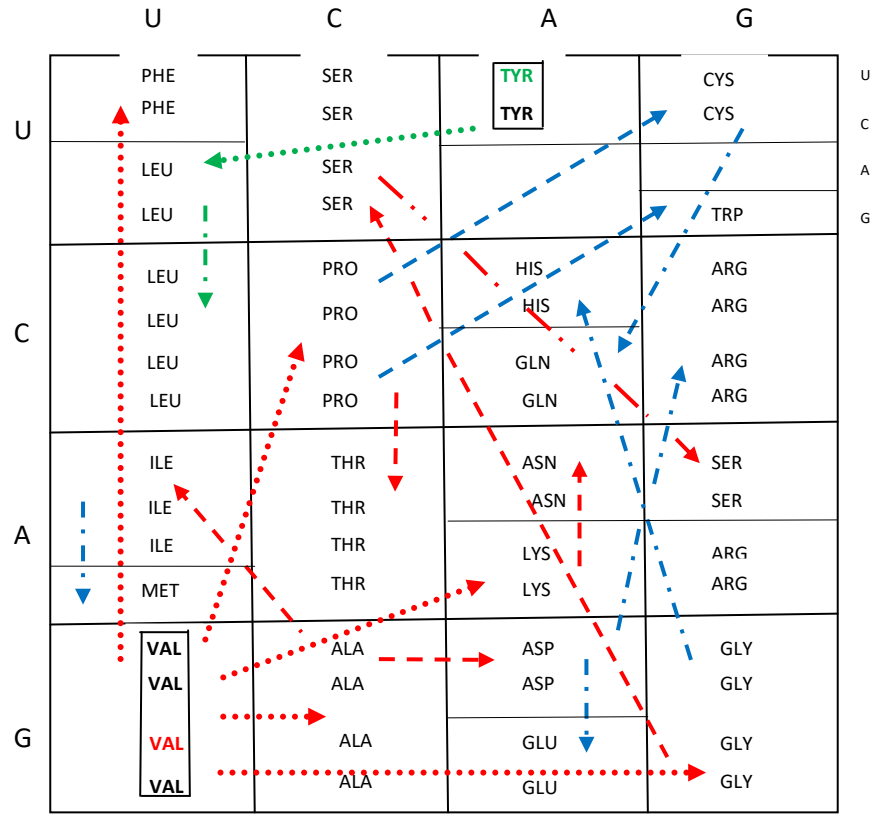

**Fig. 4**: **Synopsis of time sequence of codon adoption by amino acids**. Results are only from *Bacillus subtilis*. Red arrows: descendance from tRNA-valine into Box 1-a and b; blue arrows: descendance from Box 1 into Box 2, including serine; green arrows: descendance from tyrosine. First generation (┈┈┈►); second generation (╶╶╶►); third generation (╶·╶·►); fourth generation (━··►). For leucine, the second generation, occurring inside the lower half of the UUX square tile, is omitted. For serine, the third generation, occurring inside the UCX square tile, is omitted, as is the second generation inside the GGX square tile for glycine. See text.

Finally, it must be mentioned that the Frozen Accident Theory [3], despite its vagueness, is an interesting idea to apply to the beginning of life's evolution. In the particular case of the construction of the genetic coding apparatus as revealed by the tRNA genealogical tree, the interpretation of at least two sets of data seems to conform to this view. First, the finding that four out of five amino acids of Box1-a correspond to codons with identical bases at positions 1 and 2 is tentatively understood in reference to their corresponding helper RNA, thus making these events rather accidental and opportunistic, even while being the result of natural selection, which have definitely frozen up in the Universal Codon Table. The second example of a putative frozen accident can be seen in the late (Box 2) colonization by the remaining amino acids of the *shared* square tiles of the Universal Codon Table. Indeed, it appears that these remaining amino acids were occupying the remaining spots that were rather freely available in this table. In a previous work [9], an example of an opportunistic event was described, concerning the likely change of an amino acyl-tRNA synthetase type that would have occurred very early in evolution for the three hydrophobic amino acids valine, leucine and isoleucine. This would be tinkering at its best, in accordance with Francois Jacob's view of the way evolution has proceeded all along [20]. As strongly suggested here, many more tinkering events seem to have assisted in the formation of the Universal Codon Table, and we could maybe rename the Frozen Accident Theory as "Frozen Accident and Tinkering Theory".

**Conclusions**

The interpretation of the data presented in this work is rather straightforward since it relies on the fact that the various tRNAs of a model bacterium could be positioned into a genealogical tree by comparing their primary structure. It should be stressed that this tree does not appear to be coincidental since all of the many binary comparisons made (more than 100) are self-consistent except for only less than 2% of them [8]. The specific order found in this tree is striking and invites us to seriously consider that this tree might correctly reflect the sequence of entry of the various amino acids in the Universal Codon Table at the origin of the genetic code. This conclusion, of course, is based on the premise that the tRNA backbone as we know it today was the essential actor in the formation of the genetic code, and not some smaller molecular entity that would precede the existence of tRNAs. My main argument in favor of this view is its simplicity, taking into account that the tRNA backbone structure could have readily occurred in the RNA world [8]. Indeed, this view does not require any additional stratum of events to be articulated with the final tRNA-based genetic coding.

There exist several recently published theories on the specific steps that might have been at the origin of the genetic code [21-23]. Although imaginative and interesting, these theories, pertaining to a very complex event that has occurred from so long ago, are often compelled to depend on quite speculative reasoning. In contrast, the theory presented in this work is rather directly connected to the interpretation of experimental results, which gives it a less speculative side and, moreover, makes it easier to further test it experimentally.

If my interpretation is correct, staring from a wide and quasi-cosmological perspective, it is extraordinary that a message –embedded at least in microbes –has persisted for more than 3 billion years; this major message could enlighten us on the very first steps that created life on planet Earth (and possibly even elsewhere), and at the same time, seal the fundamental status of the tRNA molecular entity in this unique adventure.

Finally, in the usual and succinct way of describing the Central Dogma of Molecular Biology, the flow of information from DNA to protein is represented as shown in Fig. 5A. However, this is misleading in terms of chemical mechanisms because there is a huge gap between the two polynucleotides on the one end, and the protein entities on the other. Indeed, a unique coding system –which was formally deciphered more than 60 years ago but whose origin has long been totally enigmatic –was inserted to bridge the two different chemical entities. At the base of the poly-tRNA theory presented here and previously is the idea that the tRNA molecules took the principal role in shaping the Universal Codon Table, and this very early on, at we conceive of as the very beginning. Pursuing this analysis, this work presents a tentative plot of the way the Universal Codon Table could have been generated, a unique story told by nucleotide sequence comparisons of tRNAs. Thus, a preferred representation of the Central Dogma of Molecular Biology is shown in Fig. 5B, where tRNA has indeed been incorporated into the formula in its place. From this, in Fig. 5C, we can represent its reverse direction which would depict the temporal order of appearance of the other crucial macromolecular entities at the beginning of Evolution, as proposed, and where the cyclic arrows indicate the virtuous spiral taking place during the constitution of the protein- synthesis machinery.

**Figure 5:**

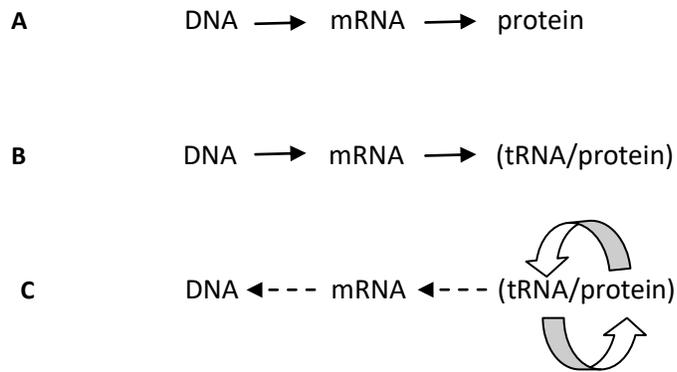

**Fig. 5**: **The Central Dogma of Molecular Biology**. **A**, The Central Dogma as information flow. **B**, The Central Dogma as successive involvement of basic macromolecular entities. **C**, Proposed time order of appearance of the macromolecular entities of the Central Dogma during primal evolution, with the autocatalytic effect of the evolving protein-synthesis machinery (arrows). See text.

**Acknowledgements:**

My thanks go to InvenTsion for sponsoring this work. Also, because theories –and the poly-tRNA theory is no exception –require strong experimental evidence for their full, or partial, eventual acceptance, I wish to thank, in advance, those scientists that would be willing to collaborate and put their expertise and techniques at the disposal of this endeavor, as well as those readers willing to add any comment to this work, in the aim of getting us closer to the truth and with a better and deeper expression. Knowing from where it comes has always been a noble goal for Humanity.